\documentclass[prb,a4paper,twocolumn,floatfix,superscriptaddress,amsmath,amsfonts,amssymb,preprintnumbers,citeautoscript,aps,longbibliography]{revtex4-2}
\pdfoutput=1 \widowpenalty10000 \clubpenalty10000
\usepackage{graphicx}
\usepackage{dcolumn}
\usepackage{bm}
\usepackage{color}
\usepackage[utf8]{inputenc}
\usepackage[T1]{fontenc}
\usepackage{booktabs}
\usepackage{microtype}
\usepackage[charter,greekuppercase=italicized]{mathdesign}
\usepackage{threeparttable}

\newcount\hh \newcount\mm
\hh=\time \divide\hh by 60
\mm=\hh \multiply\mm by 60 \mm=-\mm
\advance\mm by \time
\def\now{\number\hh:\ifnum\mm<10{}0\fi\number\mm}

\usepackage[colorlinks,plainpages=false,linkcolor=blue,urlcolor=blue,citecolor=blue,pdfpagemode=UseNone,pdfstartview=FitBH]{hyperref}

\begin{document}


\title[A.\,A.~Kulbakov \textit{et al.} Spin-waves in KCeS$_2$]{Spin-wave dynamics in the KCeS$_2$ delafossite:\\A theoretical description of powder inelastic neutron-scattering data}

\author{Stanislav~M. Avdoshenko}\email[Corresponding author: ]{s.avdoshenko@gmail.com}
\affiliation{Leibniz-Institut f\"ur Festk\"orper- und Werkstoffforschung (IFW Dresden), Helmholtzstra{\ss}e 20, 01069 Dresden, Germany.}
\author{Anton~A. Kulbakov}
 \affiliation{Institut f\"ur Festk\"orper- und Materialphysik, Technische Universit\"at Dresden, 01069 Dresden, Germany.}
\author{Ellen H\"au{\ss}ler}
\affiliation{Fakult\"at f\"ur Chemie und Lebensmittelchemie, Technische Universit\"at Dresden, 01062 Dresden, Germany.}
\author{\mbox{Philipp Schlender}}
\affiliation{Fakult\"at f\"ur Chemie und Lebensmittelchemie, Technische Universit\"at Dresden, 01062 Dresden, Germany.}
\author{Thomas Doert}
\affiliation{Fakult\"at f\"ur Chemie und Lebensmittelchemie, Technische Universit\"at Dresden, 01062 Dresden, Germany.}
\author{Jacques Ollivier}%
\affiliation{Institut Laue-Langevin, 71 avenue des Martyrs, CS 20156, 38042 Grenoble CEDEX 9, France.}
\author{Dmytro~S.~Inosov}\email[Corresponding author: \vspace{-3pt}]{dmytro.inosov@tu-dresden.de}
\affiliation{Institut f\"ur Festk\"orper- und Materialphysik, Technische Universit\"at Dresden, 01069 Dresden, Germany.}
\affiliation{W\"urzburg-Dresden Cluster of Excellence on Complexity and Topology in Quantum Matter\,---\,\textit{ct.qmat}, TU~Dresden, 01069 Dresden, Germany.}


\begin{abstract}\noindent
Layered rare-earth delafossites $A\,RX_2$ with $R$\,=\,Yb(III) or Ce(III) have received a lot of interest as potential hosts for a quantum spin-liquid ground state. Some systems of this family that show no long-range order down to the lowest measured temperatures, such as NaYbO$_2$, NaYbS$_2$, and NaYbSe$_2$, are presumed to be in a quantum spin-liquid state. However, other isostructural compounds are known to order antiferromagnetically at subkelvin temperatures. Among them, KCeS$_2$ exhibits stripe-\textit{yz} magnetic order in the triangular-lattice planes that sets in below 400~mK. Here we investigate the spin-wave spectrum of this ordered phase with powder inelastic neutron scattering and describe it using a model Hamiltonian obtained from first-principles calculations based on the complete $|J,m_J\rangle$ multiplet description of the Ce sites with an anisotropic nearest-neighbor exchange interaction. Combing the current understanding of the exchange interaction in the systems with the effects of texturing in the powder, we have been able to model the inelastic neutron scattering spectrum with high fidelity.
\end{abstract}

\maketitle
\enlargethispage{6pt}

\section{Introduction}

\begin{figure}
\includegraphics[width=\columnwidth]{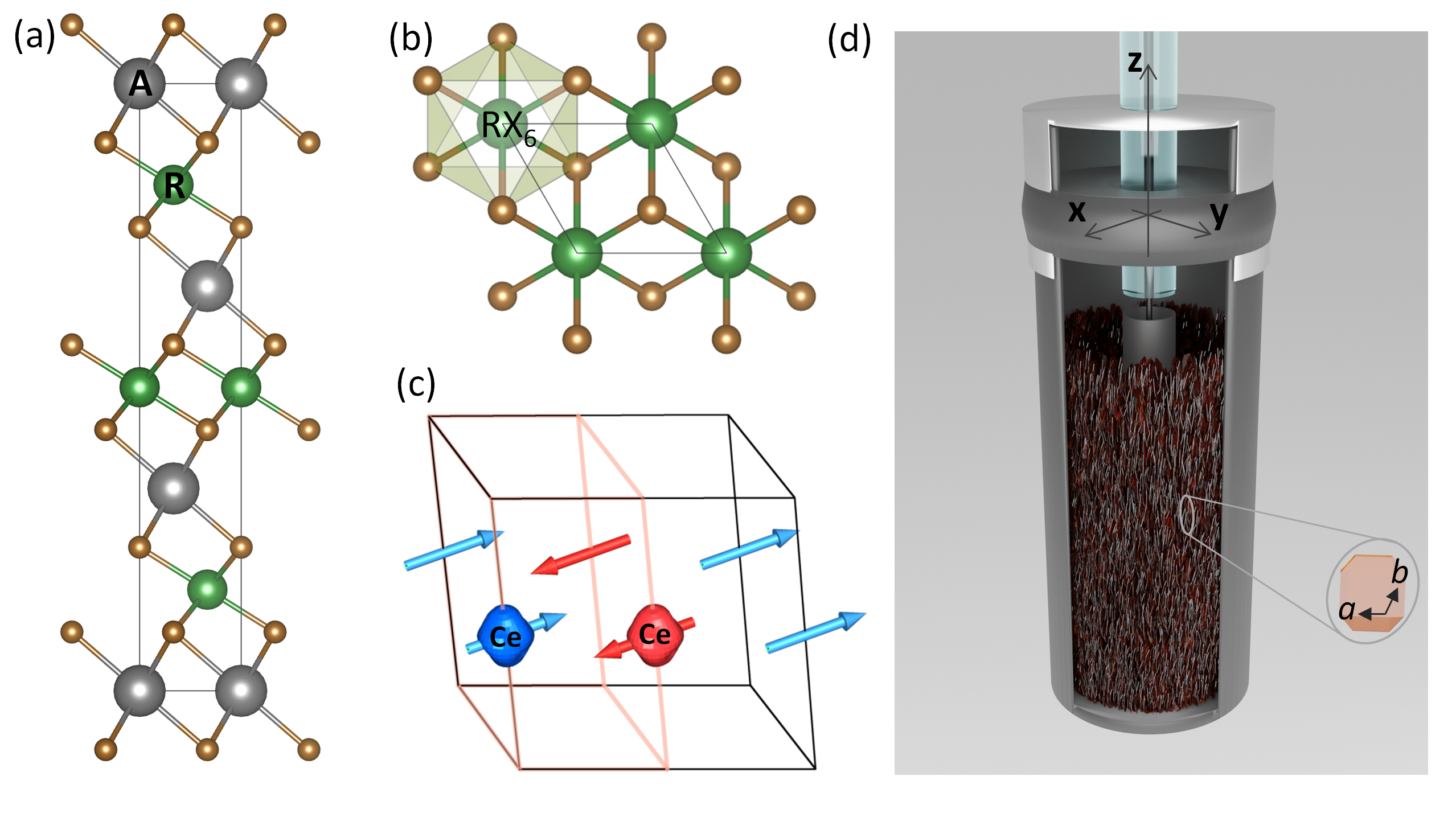}
\caption{(a)~Delafossite $ARX_2$ atomic structure in the \textit{ac}-plane with the $A^+$ and $R^{3+}$ cation sites indicated. (b)~Top view of the $R$-layer forming a triangular lattice, with the $RX_6$ polyhedron highlighted. (c)~On-site total moments and the 4\textit{f} charge density isofurface in the stripe-\textit{yz} AFM ordered state. (d)~A sketch of the powder holder used in INS experiments, with the platelike microcrystals packed so that their $c$-axes point preferentially orthogonal to the $z$-axis of the lab frame, as we found in the work (see text for details).}
\label{fig:sys_art}
\end{figure}

In the ongoing search for quantum materials, a key challenge is to find systems that exhibit strong correlations between electrons while the atomic structure supports competing interactions~\cite{Lacroix2011, BroholmCava20}. For example, triangular 2D systems are naturally anisotropic and can host strongly non-Heisenberg or compasslike interactions~\cite{NussinovBrink15, RousochatzakisRoessler16} that in certain antiferromagnetic systems are known to support quantum spin-liquid (QSL) ground states~\cite{ChaloupkaJackeli10, SinghManni12, YamajiNomura14, AlpichshevMahmood15, KitagawaTakayama18, TakayamaKato15, ModicSmidt14, SandilandsTian15, NasuKnolle16, BanerjeeBridges16, BanerjeeYan17, GordonCatuneanu19, TakagiTakayama19}. Compounds with the delafossite structure ($ARX_2$, with $A$\,--\,alkali metal, $R$\,--\,trivalent rare-earth metal, and $X$\,--\,chalcogen atom) form a family of magnetic systems with a potential to host multiple quantum phases including QSL or nontrivial magnetic orders whenever the $R$ ion is magnetic~\cite{LiuZhang18, BaenitzSchlender18, OhtaniHonjo87, KippVanderah90}. A layered hexagonal structure shown in Fig.~\ref{fig:sys_art}\,(a), with magnetic layers composed of edge-sharing $RX_6$ polyhedra [Fig.~\ref{fig:sys_art}(b)], provides an excellent platform for quantum-states design on a structurally perfect quasi-2D triangular lattice. In this research, the Ce$^{3+}$(4f$^1$) and Yb$^{3+}$(4f$^{13}$) are the ions of choice, as these Kramers cations produce a well-separated ground-state doublet (Kramers doublet, $|\text{KD}_0\rangle$) in the electrostatic potential of the $X_6$ octahedron, from which the effective spin-$\frac{1}{2}$ physics is expected.

In reality, however, the situation is more complex than that due to the mixed nature of the $J$ state in these doublets, which may undermine the spin-1/2 approximation. For example, in the case of the CeS$_6$ polyhedron in KCeS$_2$ that is shown in Fig.~\ref{fig:sys_art}(b), with $|\text{KD}_0\rangle=0.96|1/2\rangle+0.04|5/2\rangle$~\cite{Kulbakov_2021}, the magnetization density has a non-negligible anisotropy. In experiments, KCeS$_2$ develops magnetic order below $\sim$\,400~mK already in zero field, according to specific-heat measurements~\cite{Bastien2020} and neutron diffraction~\cite{Kulbakov_2021}, whereas the isostructural Yb(III)-based systems NaYbO$_2$~\cite{LiuZhang18, BordelonKenney19, DingManuel19, RanjithDmytriieva19, BordelonLiu20}, NaYbS$_2$~\cite{LiuZhang18, BaenitzSchlender18, Sarkar2019}, and NaYbSe$_2$~\cite{LiuZhang18, RanjithLuther19, ZhangLi20, DaiZhang20} show no signs of magnetic ordering down to subkelvin temperatures, thus indicating that these delafossites may have QSL ground states. In some other Yb-based systems like CsYbSe$_2$~\cite{XingSanjeewa19, XieXing22} and KYbSe$_2$~\cite{XingSanjeewa19, ScheieGhioldi21}, weak magnetic peaks in the elastic channel were observed at ultralow temperatures, suggesting short-range spin correlations that fail to form truly long-range magnetic order in the absence of a magnetic field. The true nature of these ground states as approximants of a quantum spin liquid is being actively discussed~\cite{RanjithDmytriieva19, RanjithLuther19, XingSanjeewa19, ScheieGhioldi21, XieXing22}.

The crystal structure of KCeS$_2$ is rhombohedral, described by the space group $R\overline{3}m$~\cite{PlugVerschoor76}. The low-temperature ordered state in KCeS$_2$ is established below the N\'eel temperature $T_{\rm N}\approx400$~mK~\cite{Bastien2020}, and the magnetic structure was previously determined by us in Ref.~\cite{Kulbakov_2021} as stripe-\textit{yz} AFM order with the propagation vector $(0\frac{\overline{1}}{2}\frac{1}{2})$ and spins aligned in the \textit{ab}-plane. Understanding the spin-wave dynamics in the ordered state is crucial to extract information about the local anisotropy and exchange interactions~\cite{LiWang16, MaksimovZhu19}. Ultimately, this should help us explain the differences between the magnetic ground states in rare-earth delafossites at the microscopic level, described by an effective magnetic Hamiltonian. In this Letter, we combine powder inelastic neutron scattering (INS) with \textit{ab initio} insights into the local anisotropy and interaction in the Ce-Ce pairs and formulate a model Hamiltonian of KCeS$_2$ capable of reproducing the essential features in the spin-wave spectrum.

\vspace{-2pt}\section{Experiment}\vspace{-2pt}

The phase-pure polycrystalline sample of KCeS$_2$ with a mass of 4.2~g was synthesized by the following procedure adapted from Masuda \textit{et al.}~\cite{MasudaFujino99}. Potassium carbonate (5528~mg, 40~mmol) and cerium dioxide (516.3~mg, 3~mmol) were mixed in a 13:1 molar ratio and thoroughly ground in an agate mortar under ambient conditions. The mixture was filled into a glassy carbon boat and placed in the middle of a ceramic tube in a tube furnace. After flushing the whole apparatus including a reservoir for carbon disulfide (CS$_2$) for 30~min with argon, we heated up the furnace within 3~h to the target temperature of 1050~$^\circ$C. During this time an unloaded stream of argon (2~L/h) was maintained. While dwelling for one hour, the argon stream was increased to 6~L/h and bubbled through the reservoir with liquid CS$_2$ in order to carry CS$_2$ into the hot zone of the furnace. Subsequently the furnace cooled down within 6~h to 600~$^\circ$C and after that freely without further control to room temperature under an unloaded argon stream, which was again reduced to 2~L/h. The solidified melt within the glassy carbon boat was dissolved in distilled water. Insoluble particles were filtered via paper and funnel and washed with ethanol. Carbon particles were removed by decantation. The sample was dried and stored in air. It consists of mainly small intergrown hexagonal platelets up to 10 $\mu$m in size, whereas some larger crystals with lateral dimensions up to 2~mm also occur occasionally. The sample quality was checked with powder x-ray diffraction. No additional peaks beside those of the KCeS$_2$ compound were found. As we observed that mechanical manipulations (such as pressing, grinding, etc.) fosters decomposition of KCeS$_2$ with the formation of the Ce$_2$O$_2$S phase, we avoided grinding the sample and used the as-synthesized powder in the neutron-scattering measurements to minimize the effects of surface oxidation in air. Also in the neutron-scattering measurements, this time we did not find any signatures of foreign phases, which indicates that the optimized synthesis process helped to suppress the oxysulphide impurity phase, Ce$_2$O$_2$S, which was previously reported in Ref.~\cite{Kulbakov_2021}, below the detection limit.

\begin{figure}[t]
\includegraphics[width=\columnwidth]{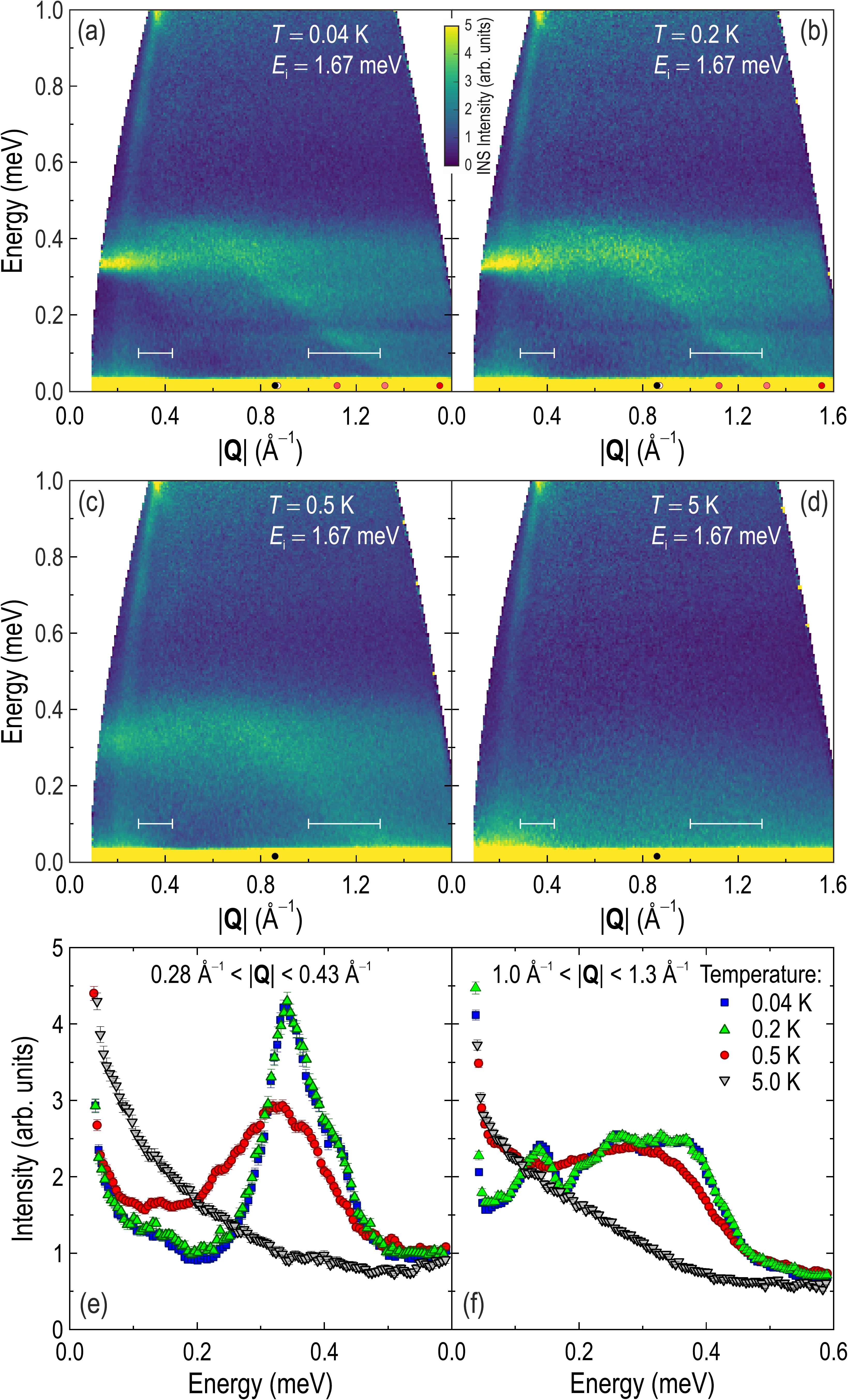}
\caption{The powder INS spectra of KCeS$_2$ at $T=0.04$, 0.2, 0.5, and 5.0~K, measured on IN5 with the incident neutron energy $E_\text{i}=1.67$~meV. (a--d)~Powder-averaged intensity maps. The black and red dots above the bottom axis show the position of the structural (003) and four magnetic Bragg peaks $\smash{(0\frac{\overline{1}}{2}\frac{1}{2})}$, $\smash{(0\frac{\overline{1}}{2}\frac{\overline{5}}{2})}$, $\smash{(0\frac{\overline{1}}{2}\frac{\overline{7}}{2})}$, and $\smash{(1\frac{\overline{1}}{2}\frac{\overline{3}}{2})}$, respectively, with their Bragg intensities encoded with color saturation. (e,\,f)~Comparison of the INS spectra that were integrated within the two momentum windows shown in the top panels with horizontal white bars.}
\label{Fig:IN5}
\end{figure}

Inelastic neutron scattering was measured using the cold-neutron disk chopper time-of-flight (TOF) spectrometer IN5 at ILL, France. To ensure proper thermal contact, the powder sample was placed in a thin-walled Cu container in the form of an annular cylinder with the inner radius of 10~mm and outer radius of 15~mm, so that it was distributed within the 2.5~mm gap between the cylinder walls and had a height of approximately 3~cm along the cylinder axis. The sample dimensions thus corresponded to the beam cross section, $30~\text{mm}\times15~\text{mm}$. The sample container was cooled down to 40~mK using a dilution refrigerator. Because of the flat, platelike shape of the crystallites in our powder, the distribution of their orientations was not uniform. The corresponding texturing of the powder had to be taken into account when analyzing the data, as described below.

The measurements were carried out with the incident neutron energy $E_\text{i}=1.67$~meV at temperatures $T=0.04$, 0.2, 0.5, and 5~K. Figure~\ref{Fig:IN5}(a--d) shows powder-averaged INS data measured at these temperatures. The data taken below $T_\text{N}$ show a sharp spin-wave mode at low $|\mathbf{Q}|$, centered at 0.34~meV in energy, with an intensity maximum near $\mathbf{Q}=0$. At higher $|\mathbf{Q}|$, this mode disperses and splits into up to three separate peaks. In Fig.~\ref{Fig:IN5}(e), the spectrum integrated within 0.28~\AA$^{-1} < |\mathbf{Q}| < 0.43$~\AA$^{-1}$ shows a shoulder on the higher-energy side of the main peak, whereas in the data integrated within 1.0~\AA$^{-1} < |\mathbf{Q}| < 1.3$~\AA$^{-1}$, presented in Fig.~\ref{Fig:IN5}(f), this peak shifts to lower energies, and one can recognize three distinct peaks in the spectrum. A flat, almost momentum-independent intensity minimum around 0.18~meV can be seen in the color maps [Fig.~\ref{Fig:IN5}(a,\,b)] and also in Fig.~\ref{Fig:IN5}(f) as a dip in the integrated spectrum. All the mentioned features broaden considerably and then disappear above $T_\text{N}$, as can be seen in Figs.~\ref{Fig:IN5}(c) and \ref{Fig:IN5}(d).

We would like to emphasize that this appearance of the powder-averaged spectrum is highly unusual in several respects. For a conventional Heisenberg antiferromagnet, one expects an intensity maximum at the gapless Goldstone magnon mode that emanates from the ordering wave vector, and not at $\mathbf{Q}=0$. This principle also holds in isostructural Yb-based delafossite compounds CsYbSe$_2$ and KYbSe$_2$, for which spin-wave spectra measured on single crystals are available~\cite{ScheieGhioldi21, XieXing22}. Furthermore, in the mentioned compounds the spin-wave dispersion goes continuously in energy from zero all the way to the top of the magnon band. We emphasize that powder-averaging such a spectrum cannot result in the spectrum we present here in Fig.~\ref{Fig:IN5} with an intensity maximum near $\mathbf{Q}=0$ and a three-peak structure at higher near $|\mathbf{Q}|$. This suggests a qualitative difference in the magnon spectrum of KCeS$_2$ in contrast to the Yb-based delafossites, and this difference is established beyond doubt even in the powder-averaged spectrum.

\begin{table}[b]\vspace{-6pt}
\caption{\label{Tab:SOCvsMH} Energies and compositions of the six lowest \textit{ab initio} SOC states in the model given by the Hamiltonian in Eq.~\eqref{Eq:Th1} for the best-fit values of the $J_{\alpha\beta}$ tensor given in Eq.~\eqref{Eq:BestFit}. The energies are given per unit cell, i.e. scaled per two Ce atoms and per six contacts within the nearest-neighbor model.\vspace{3pt}}
\centerline{
\begin{tabular}{@{}l@{~~}c@{~~~}c@{\quad~}c@{}}
\toprule
State & SOC$^\text{a}$ & Model & States composition, main component\\
      & (meV) & (meV) & \\
\midrule
$S_1$        & 0.00  & 0.00 & 0.43$\left(|+\!1/2; -1/2\rangle + |-\!1/2; +1/2\rangle\right)$\\
$D_1$        & 0.42  & 0.42 & 0.37$\left(|-\!1/2; -1/2\rangle+|+\!1/2; +1/2\rangle\right)$\\
$D^\prime_1$ & 0.43  & 0.43 & 0.38$\left(|-\!1/2; -1/2\rangle+|+\!1/2; +1/2\rangle\right)$\\
$S_2$        & 0.86  & 0.85 & 0.39$\left(|+\!1/2; -1/2\rangle+|-\!1/2; +1/2\rangle\right)$\\
$D_2\strut\!^\dagger$ & 4.43  & 3.64 & 0.11$\left(|+\!1/2; -3/2\rangle+|-\!1/2; +3/2\rangle\right)$\\
$D^\prime_2$ & 4.44  & 3.90 & 0.18$\left(|+\!1/2; -3/2\rangle+|-\!1/2; +3/2\rangle\right)$\\
\bottomrule
\end{tabular}\vspace{3pt}
}
\begin{tablenotes}
\item $^{\hspace{-3pt}\dagger}$According to the model Hamiltonian, these states may be true singlets as indicated by their composition.
\end{tablenotes}
\end{table}

\vspace{-2pt}\section{Modeling}\vspace{-2pt}

Describing the low-energy powder INS spectrum from a theoretical standpoint is more involved than similar analysis of single-crystal data. The exchange interaction model is \textit{a priori} unknown, and even when some spin model is assumed and the problem converges to parameters fitting, the fitting results may be generally less reliable than with single-crystal data. This problem is additionally aggravated if the powder sample is textured, so that the data no longer represent a uniform powder average in momentum space but have certain $\mathbf{Q}$-space directions over- or underrepresented in the average. Such texturing is expected for samples with a highly anisotropic shape of microcrystallites, and the anisotropic $\mathbf{Q}$-space sampling must be properly accounted for in the model. In our specific case of KCeS$_2$, the platelike habitus of microcrystals in the powder leads to the preferred orientation of $c$-axes in the horizontal direction, as illustrated schematically in Fig.~\ref{fig:sys_art}(d). The same habitus was observed in the refinement of our neutron diffraction data published earlier in Ref.~\cite{Kulbakov_2021}, even if the numerical value of the preferred orientation parameter may differ in our present TOF dataset due to the different shape of the sample container.

We have already combined the DFT level of theory and \textit{ab initio} modeling in an earlier publication~\cite{Kulbakov_2021} to recover the local electronic structure of the $^2F_{5/2}$ multiplet, so that it reproduces the experimental CEF excitations and their INS cross-sections. Furthermore, we have introduced an  \textit{ab initio} scheme capable of estimating exchange parameters for model interaction Hamiltonian in the $|J_1,m_{J_1}; J_2,m_{J_2}\rangle$ basis:
\begin{equation}
\hat{\mathcal H}_{M} = \hat{\mathcal H}_{\text{CF}_1} (B^k_q) + \hat{\mathcal H}_{\text{CF}_2}(B^k_q) -2 \sum^{x,y,z}_{\alpha\beta} J_{\alpha\beta} \hat{J}_{\alpha,1} \hat{J}_{\beta,2},
\label{Eq:Th1}
\end{equation}
where $\hat{\mathcal H}_{\text{CF}_1} (B^k_q)$ and $\hat{\mathcal H}_{\text{CF}_2} (B^k_q)$ are crystal-field Hamiltonians for Ce sites, and $J_{\alpha\beta}$ is the generalized coupling tensor.

\begin{figure*}
\includegraphics[width=\textwidth]{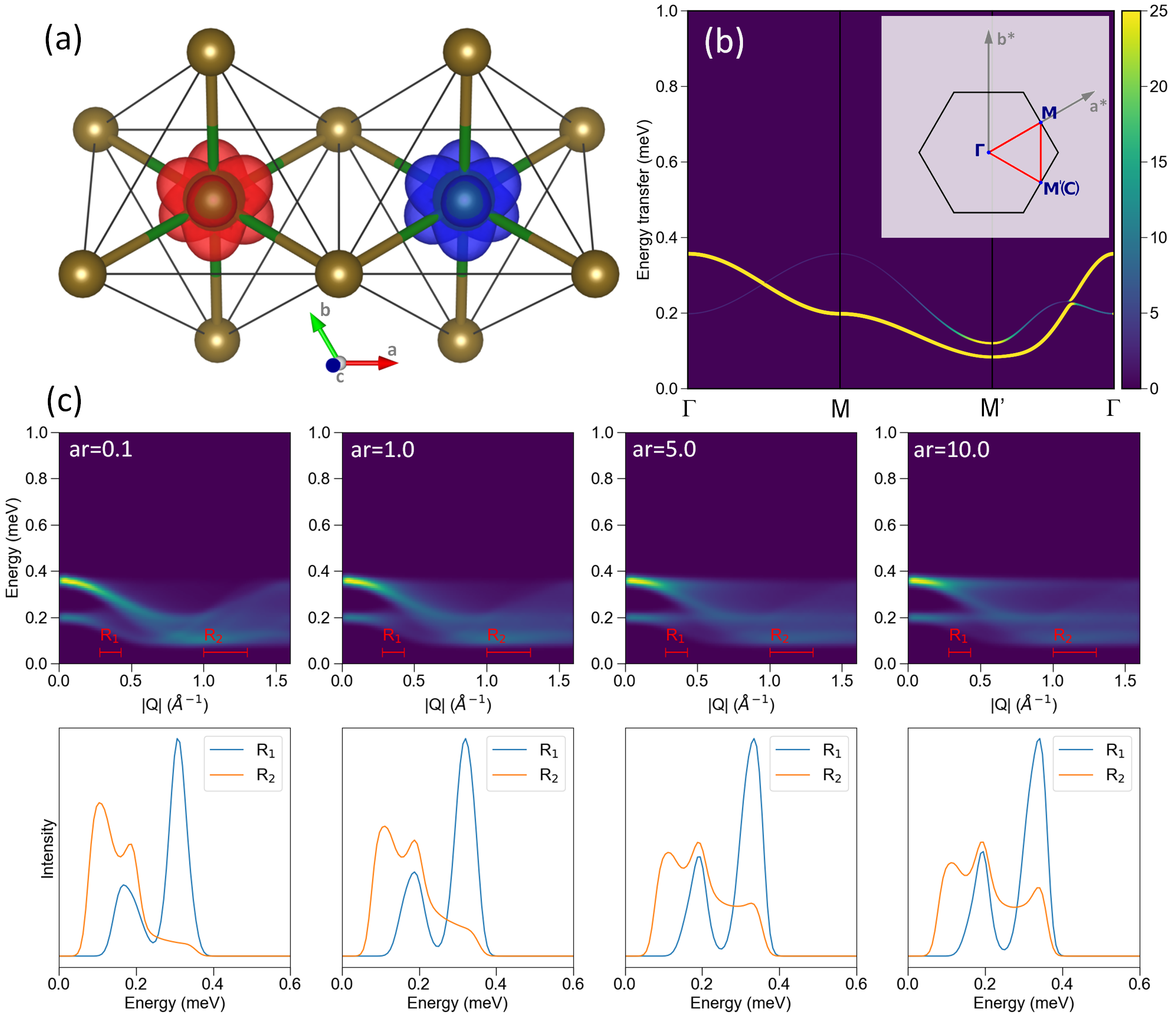}
\caption{\label{fig:model} (a)~Top view of the [Ce$_2$S$_{10}$] system considered in the dimer modeling at the CASSCF(2,14)/ANO-RCC-VDZP level, with isosurfaces of the natural spin density for the ground state. (b)~Computed dispersion of the spin waves along the high-symmetry lines ($\Gamma$--$M$--$M^\prime$--$\Gamma$), with the line thickness and color saturation proportional to $I_\text{dip}$. Here $M^\prime$ denotes the $M$ point spontaneously chosen as the magnetic ordering vector in the $ab$ plane. (c)~Theoretical INS powder spectra for the different values of aspect ratios ($\text{ar}=0.1$, 1.0, 5.0 and 10.0). The bottom row of panels shows the corresponding momentum-integrated spectra taken in the $|\mathbf{Q}|$ windows $R_1=[0.28,\,0.43]$~\AA$^{-1}$ and $R_2=[1.0,\,1.3]$~\AA$^{-1}$.}
\end{figure*}

Here we shortly recount the main steps and findings of the procedure. First, we used the dimer model for the [Ce$_2$S$_{10}$]$^{6+}$ cluster shown in Fig.~\ref{fig:model} to compute the full electronic structure with spin-orbit coupling (SOC) for CASSCF(2,14)/ANO-RCC-VDZP/RASSI/SO using \textsc{OpenMolcas} code~\cite{AquilanteAutschbach20, ChibotaruUngur12}. With system \{4f$^1$, 4f$^1$\} in active space, the reference space of the problem can be limited to AFM systems only with spin multiplicity $M=2S+1=1$ and 105 roots or combined with the FM state with $M=3$ and additional 91 roots. In this work, we optimized all 196 spin-free states and used all of them in the states interaction problem with the spin-orbit Hamiltonian. The ground-state natural spin density is shown in Fig.~\ref{fig:model}(a).  In the second step, we used Hamiltonian Eq.\ref{Eq:Th1} to fit the  $J_{\alpha\beta}$ values to reassemble the maximally close the SOC structure obtained in the \textit{ab initio} run.

When in pair, with the Ce multiplet $^2$F$_{5/2}$ and the total momentum of $J=5/2$, the Ce-Ce system has Hamiltonian dimensionality $36\times36$ on the $|J_1,m_{J_1}; J_2,m_{J_2}\rangle$ basis. Therefore, only the energies of 36 SOC states are of interest from the complete list of 378 states of the \textit{ab initio} problem. Without any constraints, the dimmer system prefers an AFM ground state with the first excited quasi-doublet within $\sim$0.4~meV followed by another singlet with excitation energy of 0.8~meV and the next quasi-double at 5.2~meV (Table \ref{Tab:SOCvsMH}). The state compositions are consistent with expectation for the AFM solution, but the states are heavily mixed in the $m_{J}$-basis.

After the best fit is established using the \textsc{PHI} program~\cite{ChiltonAnderson13}, the values of the matrix (in meV)
\begin{equation}\label{Eq:BestFit}
J_{\alpha\beta}=
\begin{pmatrix}
-0.027 &  \phantom{-}0.000  & -0.028  \\
\phantom{-}0.000 & -0.027  & \phantom{-}0.002  \\
-0.028 &  \phantom{-}0.002  & -0.160 \\
\end{pmatrix}
\end{equation}
were further used to formulate the Hamiltonian for the periodic system, taking into account the on-site $C_3$ symmetry~\cite{MaksimovZhu19}. The coupling parameters were employed in the \textsc{McPhase}~\cite{Rotter04} program for the dispersion modeling. We have reported previously, that these couplings would predict a correct order in the \textit{ab} plane~\cite{Kulbakov_2021}.

Nonetheless, the predicted energy scale for the exchange state (spin-orbit coupling $D_1$ state at 0.42~meV, Table \ref{Tab:SOCvsMH}) appears to be off by approximately 25\% compared to the experimental value (spin-wave mode energy at low $|\mathbf{Q}|$, centered around 0.34~meV in Fig.~\ref{Fig:IN5}). The need for a similar scaling level has been found in other theoretical research at the \textit{ab initio} level of theory~\cite{iwahara2015exchange}. Using the scaled \textit{J}-tensor ($J_{xx}=J_{yy}=-0.020$~meV, $J_{xz}=-0.021$~meV, $J_{zz}=-0.12$~meV)~\footnote{Mind that to use these exchanges in the \textsc{McPhase} code, values should be adjusted according to definitions in Ref.~\cite{Rotter04}}, we can recover the spin-wave dispersion along high-symmetry lines as shown in Fig.~\ref{fig:model}(b). We observe two modes with very different dipole transition moments. A minimum in the dispersion is found at the $M^\prime$ point, consistently with the in-plane projection of the ordering vector at which the Goldstone mode would be expected in the absence of spin anisotropy. The calculation predicts a significant anisotropy gap at this point and a small splitting between the two spin-wave modes due to states mixing. The higher-energy mode has vanishing intensity elsewhere in the Brillouin zone except in the vicinity of the $M^\prime$ point.

Finally, to reach better agreement of our model with the experimental powder neutron data, we have to account for the powder texturing in our sample when powder-averaging the calculated spin-wave spectrum. Because of the platelike shape of the crystallites in the powder, their orientation tends to acquire an anisotropic distribution with partially coaligned $c$-axes. Because the spectrometer is not surrounded with detectors from all sides but rather collects scattered neutrons within a horizontal slab, such an anisotropic distribution of crystallites in the powder will result in overcounting of the magnon bands from those $Q$-space directions that are preferentially oriented in the horizontal plane and undercounting of those that point preferentially in the vertical direction.

To include the effect of texturing into our model, we assume that the distribution of crystallite orientation in the powder takes the form of a spheroid, which is isotropic in the $ab$-plane and can be either oblate or prolate along the $c$-axis. Because of the cylindrically symmetric shape of the sample container (see Fig.~\ref{fig:sys_art}), it is natural to assume that the symmetry axis of the spheroid is vertical, i.e. parallel to that of the annular cylinder. This way, the texturing can be described by a single parameter\,---\,the aspect ratio ($\text{ar}$) of the ellipsoid, defined as the ratio of its $c$-axis and $a$-axis. Therefore, an isotropic powder with no texturing would correspond in this model to $\text{ar}=1$. The values $\text{ar}<1$ (an oblate spheroid) correspond to the preferentially horizontal orientation of platelike crystallites, with their $c$-axis pointing vertically along the $z$-axis in the lab frame, whereas $\text{ar}>1$ (prolate spheroid) implies that the crystallites stand on their edges and align parallel to the walls of the sample container, with the preferential orientation of the $c$-axes orthogonal to the $z$-axis. As a function of spherical angles $(\phi,\theta)$, the probability density function (PDF) representing the weights in $\mathbf{Q}$-space takes the form\vspace{-2pt}
\begin{equation*}
\text{PDF}(\theta,\text{ar})= \frac{3\,\text{ar}^{2/3}}{4\pi\sqrt{\smash{\cos}^2\theta + \text{ar}^2\,\smash{\sin}^2\theta\strut}},
\end{equation*}
where the azimuthal angle $\theta$ is measured from the $\mathbf{Q}_z$-axis. Consequently, $\text{ar}<1$ would suppress the contribution to the powder average from the regions of $\mathbf{Q}$-space with a large $Q_z$ component and enhance those from the vicinity of the $Q_xQ_y$ plane, whereas $\text{ar}>1$, on the contrary, would enhance the out-of-plane contribution and suppress the in-plane contribution.

Figure~\ref{fig:model}(c) shows the computed powder-averaged INS spectra with $\mathbf{Q}$-space sampling weights according to PDFs with $\text{ar}=0.1$, 1.0, 5.0, and 10.0. With the preferential sampling along $Q_z$ at large $\text{ar}$, a flat nondispersive feature develops in the spectrum at $\sim$\,0.38~meV. It is naturally explained by the spin-wave contributions from the vicinity of the $(0\,0\,Q_z)$ direction in reciprocal space, as they have no dispersion because of the purely 2D character of our model. The other spin-wave branch, which disperses downward from the $\Gamma$-point, originates from the in-plane contributions in the vicinity of the $Q_xQ_y$-plane in $\mathbf{Q}$-space. There is a clear resemblance of these results with the experimental spectrum in Fig.~\ref{Fig:IN5}(a), which also shows one flat feature and another dispersive peak that both emanate from $\sim$0.38~meV at the $\Gamma$ point. The best agreement is achieved for $\text{ar} \gg 1$, with $\text{ar}$ up to 10.0 being plausible when considering the highly platelike shape of the crystallites.

The intensity minimum seen in the experimental data around 0.18~meV can also be well reproduced by the model. This apparent gap in the powder-averaged spectrum appears between the flat feature extending from the weaker 0.2~meV magnon at the $\Gamma$ point due to oversampling along the $c$-axis and a low-energy peak that originates from the bottom of the magnon band at $\sim$0.1~meV near the $M^\prime$-point.

The less intense 0.2~meV magnon at the $\Gamma$ point, suggested by our calculations, is not observed in the experimental data at low $|\mathbf{Q}|$, and it is open for interpretation of the experimental INS. The integrated INS spectra Fig.~\ref{Fig:IN5}(e) within the window at small $|\mathbf{Q}|$ has a signal at these energies. Although, admittedly, theory predicts a much stronger signal in relation to the main feature at 0.35~meV, as one can see in the integrated spectra in Fig.~\ref{fig:model}(c). Nevertheless, these allow the conclusion that our model recovers all pronounced features of experimental spectra and explains its qualitative differences from the magnon spectra of Yb-based delafossite materials. Some quantitative differences between the experimental data and the model can be due to the fact that we constrained our model to a single two-dimensional layer with only nearest-neighbor interactions and used a simple single-parameter approximation to describe the preferential domain distribution in the powder sample, which in reality can be more complex.

In summary, we presented a detailed study of the spin-wave spectrum in the magnetically ordered phase of KCeS$_\text{2}$ with powder inelastic neutron scattering. Based on well-resolved experimental data and sophisticated theoretical modeling supported by \textit{ab initio} derived exchange interactions and powder texturing effects, it was possible to rationalize the observed anomalies in the INS spectra and confirm that the spin model of KCeS$_\text{2}$ involves strongly anisotropic non-Heisenberg nearest-neighbor exchange interactions with $J_{zz} \gg J_{xx}\!= J_{yy}$.

\vspace{-2pt}\section*{Acknowledgments}\vspace{-2pt}

The authors are very thankful to S.~Nikitin from Paul Scherrer Institute, Switzerland, and to T.~Xie and A.~Podlesnyak from Oak Ridge National Laboratory for discussions and sharing their unpublished data. S.M.A. also thanks M.~Vojta (TU Dresden) for fruitful discussions. This project was funded in part by the German Research Foundation (DFG) under the individual research Grants \mbox{AV 169/3-1}, \mbox{IN 209/9-1}, via the projects B03 and C03 of the Collaborative Research Center SFB~1143 (project-id 247310070) at the TU Dresden, and the W\"urzburg-Dresden Cluster of Excellence on Complexity and Topology in Quantum Matter\,---\,\textit{ct.qmat} (EXC 2147, project-id 390858490).

\bibliographystyle{my-apsrev}
\bibliography{KCeS2_SpinWaves}

\end{document}